\input harvmac
\let\includefigures=\iftrue
\let\useblackboard==\iftrue
\newfam\black

\includefigures

\let\includefigures=\iftrue
\let\useblackboard=\iftrue
\newfam\black

\includefigures
\message{If you do not have epsf.tex (to include figures),}
\message{change the option at the top of the tex file.}
\input epsf
\def\figin{\epsfcheck\figin}\def\figins{\epsfcheck\figins}
\def\epsfcheck{\ifx\epsfbox\UnDeFiNeD
\message{(NO epsf.tex, FIGURES WILL BE IGNORED)}
\gdef\figin##1{\vskip2in}\gdef\figins##1{\hskip.5in}
\else\message{(FIGURES WILL BE INCLUDED)}%
\gdef\figin##1{##1}\gdef\figins##1{##1}\fi}
\def\DefWarn#1{}
\def\figinsert{\goodbreak\midinsert}
\def\ifig#1#2#3{\DefWarn#1\xdef#1{fig.~\the\figno}
\writedef{#1\leftbracket fig.\noexpand~\the\figno}%
\figinsert\figin{\centerline{#3}}\medskip\centerline{\vbox{
\baselineskip12pt\advance\hsize by -1truein
\noindent\footnotefont{\bf Fig.~\the\figno:} #2}}
\endinsert\global\advance\figno by1}
\else
\def\ifig#1#2#3{\xdef#1{fig.~\the\figno}
\writedef{#1\leftbracket fig.\noexpand~\the\figno}%
\global\advance\figno by1} \fi

\def\journal#1&#2(#3){\unskip, \sl #1\ \bf #2 \rm(19#3) }
\def\andjournal#1&#2(#3){\sl #1~\bf #2 \rm (19#3) }

\def\ie{{\it i.e.}}

\noblackbox
%


\def\unlockat{\catcode`\@=11}
\def\lockat{\catcode`\@=12}

\unlockat

\def\newsec#1{\global\advance\secno by1\message{(\the\secno. #1)}
\global\subsecno=0\global\subsubsecno=0\eqnres@t\noindent
{\bf\the\secno. #1}
\writetoca{{\secsym} {#1}}\par\nobreak\medskip\nobreak}
\global\newcount\subsecno \global\subsecno=0
\def\subsec#1{\global\advance\subsecno
by1\message{(\secsym\the\subsecno. #1)}
\ifnum\lastpenalty>9000\else\bigbreak\fi\global\subsubsecno=0
\noindent{\it\secsym\the\subsecno. #1}
\writetoca{\string\quad {\secsym\the\subsecno.} {#1}}
\par\nobreak\medskip\nobreak}
\global\newcount\subsubsecno \global\subsubsecno=0
\def\subsubsec#1{\global\advance\subsubsecno by1
\message{(\secsym\the\subsecno.\the\subsubsecno. #1)}
\ifnum\lastpenalty>9000\else\bigbreak\fi
\noindent\quad{\secsym\the\subsecno.\the\subsubsecno.}{#1}
\writetoca{\string\qquad{\secsym\the\subsecno.\the\subsubsecno.}{#1}}
\par\nobreak\medskip\nobreak}

\def\subsubseclab#1{\DefWarn#1\xdef
#1{\noexpand\hyperref{}{subsubsection}%
{\secsym\the\subsecno.\the\subsubsecno}%
{\secsym\the\subsecno.\the\subsubsecno}}%
\writedef{#1\leftbracket#1}\wrlabeL{#1=#1}}
\lockat

\def\ie{{\it i.e.}}


\font\manual=manfnt \def\dbend{\lower3.5pt\hbox{\manual\char127}}

\def\IZ{\relax\ifmmode\mathchoice
{\hbox{\cmss Z\kern-.4em Z}}{\hbox{\cmss Z\kern-.4em Z}}
{\lower.9pt\hbox{\cmsss Z\kern-.4em Z}}
{\lower1.2pt\hbox{\cmsss Z\kern-.4em Z}}\else{\cmss Z\kern-.4em
Z}\fi}


\def\IZ{\relax\ifmmode\mathchoice
{\hbox{\cmss Z\kern-.4em Z}}{\hbox{\cmss Z\kern-.4em Z}}
{\lower.9pt\hbox{\cmsss Z\kern-.4em Z}}
{\lower1.2pt\hbox{\cmsss Z\kern-.4em Z}}\else{\cmss Z\kern-.4em
Z}\fi}
\def\IB{\relax{\rm I\kern-.18em B}}
\def\IC{{\relax\hbox{$\inbar\kern-.3em{\rm C}$}}}
\def\ID{\relax{\rm I\kern-.18em D}}
\def\IE{\relax{\rm I\kern-.18em E}}
\def\IF{\relax{\rm I\kern-.18em F}}
\def\IG{\relax\hbox{$\inbar\kern-.3em{\rm G}$}}
\def\IGa{\relax\hbox{${\rm I}\kern-.18em\Gamma$}}
\def\IH{\relax{\rm I\kern-.18em H}}
\def\II{\relax{\rm I\kern-.18em I}}
\def\IK{\relax{\rm I\kern-.18em K}}
\def\IP{\relax{\rm I\kern-.18em P}}
\def\IQ{\relax\hbox{$\inbar\kern-.3em{\rm Q}$}}

\def\inbar{\,\vrule height1.5ex width.4pt depth0pt}

\font\cmss=cmss10 \font\cmsss=cmss10 at 7pt
\def\IR{\relax{\rm I\kern-.18em R}}

%
%

\def\makeblankbox#1#2{\hbox{\lower\dp0\vbox{\hidehrule{#1}{#2}%
   \kern -#1
   \hbox to \wd0{\hidevrule{#1}{#2}%
      \raise\ht0\vbox to #1{}
      \lower\dp0\vtop to #1{}
      \hfil\hidevrule{#2}{#1}}%
   \kern-#1\hidehrule{#2}{#1}}}%
}%
\def\hidehrule#1#2{\kern-#1\hrule height#1 depth#2 \kern-#2}%
\def\hidevrule#1#2{\kern-#1{\dimen0=#1\advance\dimen0 by #2\vrule
    width\dimen0}\kern-#2}%
\def\openbox{\ht0=1.2mm \dp0=1.2mm \wd0=2.4mm  \raise 2.75pt
\makeblankbox {.25pt} {.25pt}  }

\def\bun#1/#2{\leavevmode
   \kern.1em \raise .5ex \hbox{\the\scriptfont0 #1}%
   \kern-.1em $/$%
   \kern-.15em \lower .25ex \hbox{\the\scriptfont0 #2}%
}

\def\opensquare{\ht0=3.4mm \dp0=3.4mm \wd0=6.8mm  \raise 2.7pt
\makeblankbox {.25pt} {.25pt}  }


\def\sector#1#2{\ {\scriptstyle #1}\hskip 1mm
\mathop{\opensquare}\limits_{\lower 1mm\hbox{$\scriptstyle#2$}}\hskip 1mm}

\def\tsector#1#2{\ {\scriptstyle #1}\hskip 1mm
\mathop{\opensquare}\limits_{\lower 1mm\hbox{$\scriptstyle#2$}}^\sim\hskip 1mm}


\def\inbar{\,\vrule height1.5ex width.4pt depth0pt}

\font\cmss=cmss10 \font\cmsss=cmss10 at 7pt
\def\IR{\relax{\rm I\kern-.18em R}}


\def\frac#1#2{{#1\over#2}}

\def\inbar{\,\vrule height1.5ex width.4pt depth0pt}
\def\IC{\relax\hbox{$\inbar\kern-.3em{\rm C}$}}
\def\IR{\relax{\rm I\kern-.18em R}}
\def\IP{\relax{\rm I\kern-.18em P}}

%
%
\catcode`\@=11
\def\slash#1{\mathord{\mathpalette\c@ncel{#1}}}
\overfullrule=0pt

\def\II{{\cal I}}

\def\underrel#1\over#2{\mathrel{\mathop{\kern\z@#1}\limits_{#2}}}

\catcode`\@=12


%

\def\sinh{{\rm sinh}}
\def\cosh{{\rm cosh}}

\def\exp{{\rm exp}}



\def\frac#1#2{{#1\over#2}}

\def\inbar{\,\vrule height1.5ex width.4pt depth0pt}
\def\IC{\relax\hbox{$\inbar\kern-.3em{\rm C}$}}
\def\IR{\relax{\rm I\kern-.18em R}}
\def\IP{\relax{\rm I\kern-.18em P}}

%
%

%
\catcode`\@=11
\def\slash#1{\mathord{\mathpalette\c@ncel{#1}}}
\overfullrule=0pt

\def\II{{\cal I}}

\def\underrel#1\over#2{\mathrel{\mathop{\kern\z@#1}\limits_{#2}}}

\catcode`\@=12


%

\def \sinh{{\rm sinh}}
\def \cosh{{\rm cosh}}

\def\exp{{\rm exp}}


\lref\WakimotoGF{
  M.~Wakimoto,
  ``Fock representations of the affine lie algebra A1(1),''
Commun.\ Math.\ Phys.\  {\bf 104}, 605 (1986).
}

\lref\KazakovPM{
  V.~Kazakov, I.~K.~Kostov and D.~Kutasov,
  ``A Matrix model for the two-dimensional black hole,''
Nucl.\ Phys.\ B {\bf 622}, 141 (2002).
[hep-th/0101011].
}
\lref\BershadskyMF{
  M.~Bershadsky and H.~Ooguri,
  ``Hidden SL(n) Symmetry in Conformal Field Theories,''
Commun.\ Math.\ Phys.\  {\bf 126}, 49 (1989).
}

\lref\AharonyXN{
  O.~Aharony, A.~Giveon and D.~Kutasov,
  ``LSZ in LST,''
Nucl.\ Phys.\ B {\bf 691}, 3 (2004).
[hep-th/0404016].
}

\lref\PolchinskiTA{
  J.~Polchinski,
  ``String theory and black hole complementarity,''
[hep-th/9507094].
}

\lref\GerasimovFI{
  A.~Gerasimov, A.~Morozov, M.~Olshanetsky, A.~Marshakov and S.~L.~Shatashvili,
  ``Wess-Zumino-Witten model as a theory of free fields,''
Int.\ J.\ Mod.\ Phys.\ A {\bf 5}, 2495 (1990).
}

\lref\ItzhakiJT{
  N.~Itzhaki,
  ``Is the black hole complementarity principle really necessary?,''
[hep-th/9607028].
}

\lref\BraunsteinMY{
  S.~L.~Braunstein, S.~Pirandola and K.~Życzkowski,
  ``Better Late than Never: Information Retrieval from Black Holes,''
Phys.\ Rev.\ Lett.\  {\bf 110}, no. 10, 101301 (2013).
[arXiv:0907.1190 [quant-ph]].
}

\lref\WittenZW{
  E.~Witten,
  ``Anti-de Sitter space, thermal phase transition, and confinement in gauge theories,''
Adv.\ Theor.\ Math.\ Phys.\  {\bf 2}, 505 (1998).
[hep-th/9803131].
}

\lref\GoulianQR{
  M.~Goulian and M.~Li,
  ``Correlation functions in Liouville theory,''
Phys.\ Rev.\ Lett.\  {\bf 66}, 2051 (1991).
}

\lref\AlmheiriRT{
  A.~Almheiri, D.~Marolf, J.~Polchinski and J.~Sully,
  ``Black Holes: Complementarity or Firewalls?,''
JHEP {\bf 1302}, 062 (2013).
[arXiv:1207.3123 [hep-th]].
}

\lref\MarolfDBA{
  D.~Marolf and J.~Polchinski,
  ``Gauge/Gravity Duality and the Black Hole Interior,''
Phys.\ Rev.\ Lett.\  {\bf 111}, 171301 (2013).
[arXiv:1307.4706 [hep-th]].
}

\lref\PolchinskiCEA{
  J.~Polchinski,
  ``Chaos in the black hole S-matrix,''
[arXiv:1505.08108 [hep-th]].
}

\lref\MathurHF{
  S.~D.~Mathur,
  ``The Information paradox: A Pedagogical introduction,''
Class.\ Quant.\ Grav.\  {\bf 26}, 224001 (2009).
[arXiv:0909.1038 [hep-th]].
}

\lref\BershadskyIN{
  M.~Bershadsky and D.~Kutasov,
  ``Comment on gauged WZW theory,''
Phys.\ Lett.\ B {\bf 266}, 345 (1991).
}

\lref\GiveonCMA{
  A.~Giveon, N.~Itzhaki and D.~Kutasov,
  ``Stringy Horizons,''
JHEP {\bf 1506}, 064 (2015).
[arXiv:1502.03633 [hep-th]].
}

\lref\HartleAI{
  J.~B.~Hartle and S.~W.~Hawking,
  ``Wave Function of the Universe,''
Phys.\ Rev.\ D {\bf 28}, 2960 (1983), [Adv.\ Ser.\ Astrophys.\ Cosmol.\  {\bf 3}, 174 (1987)].
}

\lref\GrossKZA{
  D.~J.~Gross and P.~F.~Mende,
  ``The High-Energy Behavior of String Scattering Amplitudes,''
Phys.\ Lett.\ B {\bf 197}, 129 (1987).
}
\lref\GrossAR{
  D.~J.~Gross and P.~F.~Mende,
  ``String Theory Beyond the Planck Scale,''
Nucl.\ Phys.\ B {\bf 303}, 407 (1988).
}

\lref\ItzhakiGLF{
  N.~Itzhaki,
  ``Stringy instability inside the black hole,''
JHEP {\bf 1810}, 145 (2018).
[arXiv:1808.02259 [hep-th]].
}

\lref\ItzhakiRLD{
  N.~Itzhaki and L.~Liram,
  ``A stringy glimpse into the black hole horizon,''
JHEP {\bf 1804}, 018 (2018).
[arXiv:1801.04939 [hep-th]].
}

\lref\Ben{
  R.~Ben-Israel, A.~Giveon, N.~Itzhaki and L.~Liram,
  ``On the black hole interior in string theory,''
JHEP {\bf 1705}, 094 (2017).
[arXiv:1702.03583 [hep-th]].
}

\lref\PolyakovVU{
  A.~M.~Polyakov,
  ``Thermal Properties of Gauge Fields and Quark Liberation,''
Phys.\ Lett.\  {\bf 72B}, 477 (1978).
}

\lref\SusskindUP{
  L.~Susskind,
  ``Lattice Models of Quark Confinement at High Temperature,''
Phys.\ Rev.\ D {\bf 20}, 2610 (1979).
}

\lref\KutasovXU{
  D.~Kutasov and N.~Seiberg,
  ``More comments on string theory on AdS(3),''
JHEP {\bf 9904}, 008 (1999).
[hep-th/9903219].
}

\lref\TeschnerUG{
  J.~Teschner,
  ``Operator product expansion and factorization in the H+(3) WZNW model,''
Nucl.\ Phys.\ B {\bf 571}, 555 (2000).
[hep-th/9906215].
}

\lref\GiribetFT{
  G.~Giribet and C.~A.~Nunez,
  ``Correlators in AdS(3) string theory,''
JHEP {\bf 0106}, 010 (2001).
[hep-th/0105200].
}

\lref\MaldacenaRE{
  J.~M.~Maldacena,
  ``The Large N limit of superconformal field theories and supergravity,''
Int.\ J.\ Theor.\ Phys.\  {\bf 38}, 1113 (1999), [Adv.\ Theor.\ Math.\ Phys.\  {\bf 2}, 231 (1998)].
[hep-th/9711200].
}

\lref\MaldacenaHW{
  J.~M.~Maldacena and H.~Ooguri,
  ``Strings in AdS(3) and SL(2,R) WZW model 1.: The Spectrum,''
J.\ Math.\ Phys.\  {\bf 42}, 2929 (2001).
[hep-th/0001053].
}

\lref\DiFrancescoOCM{
  P.~Di Francesco and D.~Kutasov,
  ``Correlation functions in 2-D string theory,''
Phys.\ Lett.\ B {\bf 261}, 385 (1991).
}

\lref\DiFrancescoDAF{
  P.~Di Francesco and D.~Kutasov,
  ``World sheet and space-time physics in two-dimensional (Super)string theory,''
Nucl.\ Phys.\ B {\bf 375}, 119 (1992).
[hep-th/9109005].
}

\lref\MaldacenaKM{
  J.~M.~Maldacena and H.~Ooguri,
  ``Strings in AdS(3) and the SL(2,R) WZW model. Part 3. Correlation functions,''
Phys.\ Rev.\ D {\bf 65}, 106006 (2002).
[hep-th/0111180].
}

\lref\MaldacenaKY{
  J.~M.~Maldacena,
  ``Black holes in string theory,''
[hep-th/9607235].
}

\lref\HawkingSW{
  S.~W.~Hawking,
  ``Particle Creation by Black Holes,''
Commun.\ Math.\ Phys.\  {\bf 43}, 199 (1975), Erratum: [Commun.\ Math.\ Phys.\  {\bf 46}, 206 (1976)].
}

\lref\PeetHN{
  A.~W.~Peet,
  ``TASI lectures on black holes in string theory,''
[hep-th/0008241].
}

\lref\GiveonDXE{
  A.~Giveon, N.~Itzhaki and D.~Kutasov,
  ``Stringy Horizons II,''
JHEP {\bf 1610}, 157 (2016).
[arXiv:1603.05822 [hep-th]].
}

\lref\GiveonICA{
  A.~Giveon and N.~Itzhaki,
  ``String theory at the tip of the cigar,''
JHEP {\bf 1309}, 079 (2013).
[arXiv:1305.4799 [hep-th]].
}
\lref\AttaliGOQ{
  K.~Attali and N.~Itzhaki,
  ``The Averaged Null Energy Condition and the Black Hole Interior in String Theory,''
Nucl.\ Phys.\ B {\bf 943}, 114631 (2019).
[arXiv:1811.12117 [hep-th]].
}

\lref\MaldacenaHI{
  J.~M.~Maldacena,
  ``Long strings in two dimensional string theory and non-singlets in the matrix model,''
JHEP {\bf 0509}, 078 (2005), [Int.\ J.\ Geom.\ Meth.\ Mod.\ Phys.\  {\bf 3}, 1 (2006)].
[hep-th/0503112].
}

\lref\AharonyQU{
  O.~Aharony and E.~Witten,
  ``Anti-de Sitter space and the center of the gauge group,''
JHEP {\bf 9811}, 018 (1998).
[hep-th/9807205].
}

\lref\GiveonUP{
  A.~Giveon and D.~Kutasov,
  ``Notes on AdS(3),''
Nucl.\ Phys.\ B {\bf 621}, 303 (2002).
[hep-th/0106004].
}

\lref\KutasovRR{
  D.~Kutasov,
  ``Accelerating branes and the string/black hole transition,''
[hep-th/0509170].
}
\lref\AharonyAN{
  O.~Aharony and D.~Kutasov,
  ``Holographic Duals of Long Open Strings,''
Phys.\ Rev.\ D {\bf 78}, 026005 (2008).
[arXiv:0803.3547 [hep-th]].
}

\lref\GiribetFY{
  G.~Giribet and C.~A.~Nunez,
  ``Aspects of the free field description of string theory on AdS(3),''
JHEP {\bf 0006}, 033 (2000).
[hep-th/0006070].
}

\lref\fzz{ V.A. Fateev, A.B. Zamolodchikov and Al.B. Zamolodchikov, unpublished.
}

\lref\GiribetKCA{
  G.~Giribet and A.~Ranjbar,
  ``Screening Stringy Horizons,''
Eur.\ Phys.\ J.\ C {\bf 75}, no. 10, 490 (2015).
[arXiv:1504.05044 [hep-th]].
}

\lref\CallanIA{
  C.~G.~Callan, Jr., E.~J.~Martinec, M.~J.~Perry and D.~Friedan,
  ``Strings in Background Fields,''
Nucl.\ Phys.\ B {\bf 262}, 593 (1985).
}

\lref\AtickSI{
  J.~J.~Atick and E.~Witten,
  ``The Hagedorn Transition and the Number of Degrees of Freedom of String Theory,''
Nucl.\ Phys.\ B {\bf 310}, 291 (1988).
}

\lref\ElitzurCB{
  S.~Elitzur, A.~Forge and E.~Rabinovici,
  ``Some global aspects of string compactifications,''
Nucl.\ Phys.\ B {\bf 359}, 581 (1991).
}

\lref\MandalTZ{
  G.~Mandal, A.~M.~Sengupta and S.~R.~Wadia,
  ``Classical solutions of two-dimensional string theory,''
Mod.\ Phys.\ Lett.\ A {\bf 6}, 1685 (1991).
}

\lref\WittenYR{
  E.~Witten,
  ``On string theory and black holes,''
Phys.\ Rev.\ D {\bf 44}, 314 (1991).
}

\lref\DijkgraafBA{
  R.~Dijkgraaf, H.~L.~Verlinde and E.~P.~Verlinde,
  ``String propagation in a black hole geometry,''
Nucl.\ Phys.\ B {\bf 371}, 269 (1992).
}

\lref\KutasovRR{
  D.~Kutasov,
  ``Accelerating branes and the string/black hole transition,''
[hep-th/0509170].
}

\lref\KarczmarekBW{
  J.~L.~Karczmarek, J.~M.~Maldacena and A.~Strominger,
  ``Black hole non-formation in the matrix model,''
JHEP {\bf 0601}, 039 (2006).
[hep-th/0411174].
}

\lref\BenIsraelMDA{
  R.~Ben-Israel, A.~Giveon, N.~Itzhaki and L.~Liram,
  ``Stringy Horizons and UV/IR Mixing,''
JHEP {\bf 1511}, 164 (2015).
[arXiv:1506.07323 [hep-th]].
}

\lref\MaldacenaCG{
  J.~M.~Maldacena and A.~Strominger,
  ``Semiclassical decay of near extremal five-branes,''
JHEP {\bf 9712}, 008 (1997).
[hep-th/9710014].
}

\lref\GiveonMI{
  A.~Giveon, D.~Kutasov, E.~Rabinovici and A.~Sever,
  ``Phases of quantum gravity in AdS(3) and linear dilaton backgrounds,''
Nucl.\ Phys.\ B {\bf 719}, 3 (2005).
[hep-th/0503121].
}

\lref\BarsRB{
  I.~Bars and D.~Nemeschansky,
  ``String Propagation in Backgrounds With Curved Space-time,''
Nucl.\ Phys.\ B {\bf 348}, 89 (1991).
}

\lref\BarsSV{
  I.~Bars and J.~Schulze,
  ``Folded strings falling into a black hole,''
Phys.\ Rev.\ D {\bf 51}, 1854 (1995).
[hep-th/9405156].
}

\lref\BarsXI{
  I.~Bars,
  ``Folded strings in curved space-time,''
[hep-th/9411078].
}

\lref\BarsQM{
  I.~Bars,
  ``Folded strings,''
Lect.\ Notes Phys.\  {\bf 447}, 26 (1995).
[hep-th/9412044].
}

\Title{
} {\vbox{
\bigskip\centerline{Stringy Black Hole Interiors}}}
\medskip
\centerline{\it Amit Giveon${}^{1}$ and Nissan Itzhaki${}^{2}$ }
\bigskip
\smallskip
\centerline{${}^{1}$Racah Institute of Physics, The Hebrew
University} \centerline{Jerusalem 91904, Israel}
\smallskip
\centerline{${}^{2}$ Physics Department, Tel-Aviv University, Israel} \centerline{Ramat-Aviv, 69978, Israel}
\smallskip

\smallskip

\vglue .3cm

\bigskip

\bigskip
\noindent

It is well known  that non-perturbative $\alpha'$ corrections to the $SL(2,\IR)/U(1)$ cigar geometry are described via a condensation of a Sine-Liouville operator that schematically can be written as $W^{+}+W^{-}$, where $W^{\pm}$ describe a string with winding number $\pm 1$. This condensation leads to  interesting effects in the cigar geometry that take place already at the classical level in string theory. Condensation of the analytically continued Sine-Liouville operator in the Lorentzian $SL(2,\IR)/U(1)$ black hole is problematic.
Here, we propose that in the black hole case,
the non-perturbative $\alpha'$ corrections are described in terms of an operator that can be viewed as the analytic continuation of the fusion of $W^+$ and $W^-$. We show that this operator does not suffer from the same problem as the analytically continued Sine-Liouville operator and argue that it describes folded strings that fill the entire black hole and, in a sense, replace the black hole interior.
We  estimate the folded strings radiation, and show that they radiate at the  Hawking temperature.

\bigskip

\Date{}

\newsec{Introduction}

Non-perturbative $\alpha'$ corrections to the $SL(2,\IR)/U(1)$ cigar geometry \refs{\BarsRB\ElitzurCB\MandalTZ\WittenYR-\DijkgraafBA}
are described \refs{\fzz\KazakovPM\GiveonUP\AharonyXN\KarczmarekBW\GiveonMI-\KutasovRR}
in terms of a condensation of a Sine-Liouville operator,
$\lambda_W (W^+ +W^- )$, where
\eqn\aaaa{W^{\pm}=\exp\left(\pm i {\beta\over 2\pi\alpha{'}}(x_L-x_R)\right) e^{-{1\over Q}\phi}~;}
$\phi$ is the linear dilaton direction, with a slope $Q$,
$x=x_L+x_R$ is the Euclidean time direction, with periodicity $\beta$,~\foot{In
the superstring, \aaaa\ is an $N=2$ Liouville superfield, with $\beta={2\pi\over Q}\sqrt{2\alpha'}$,
while in the bosonic string $\beta={2\pi\over Q}\sqrt{2(1+Q^2)\alpha'}$.}
and $\lambda_W$ is the size of the winding condensate.
These stringy corrections, that are referred to as the FZZ duality \fzz\ (see also \KazakovPM),
led \refs{\GiveonCMA,\BenIsraelMDA,\GiveonDXE} to intriguing effects,
already in classical string theory on the cigar.

The $SL(2,\IR)/U(1)$ black hole (BH) geometry can be obtained from the cigar geometry via an analytic continuation. This coset CFT is particularly interesting in string theory, e.g. since it is obtained in the near horizon limit of $k$ near extremal NS5-branes \MaldacenaCG\ (with $Q^2=2/k$). It would be very helpful to understand the non-perturbative $\alpha'$ corrections to near extremal NS5-branes; these could possibly teach us some general lessons about horizons and BH interiors  in string theory.
Naively, we simply have to Wick rotate \aaaa, to obtain
\eqn\bbbb{W^{\pm}=\exp\left(\pm {\beta\over 2\pi\alpha'}(t_L-t_R)\right) e^{-{1\over Q}\phi}~,}
where $t=t_L+t_R$ is the real time direction ($t=ix$), but
a subtlety with the operators \bbbb\ is that they are not mutually local with vertex operator
that amount to energy eigenstates.

A seemingly orthogonal question about this model was raised recently;
it is related to the argument \ItzhakiGLF\
that folded strings~\refs{\BarsSV,\BarsXI,\BarsQM} are created classically inside the $SL(2,\IR)/U(1)$ BH.
If indeed there are classical folded strings inside the BH,
then what is the operator in the $SL(2,\IR)/U(1)$ coset CFT that describes them?

The aim of this note is to address these questions.
We recall \refs{\BershadskyMF\GerasimovFI\GiribetFY-\GiribetFT} that, in addition to \bbbb,
there is another non-perturbative (1,1) operator, denoted by $F$, whose underlying $AdS_3$ parent
is invariant under the $SL(2,\IR)_L\times SL(2,\IR)_R$ current algebra;
thus, it condenses. We argue that it amounts to the BH filling folded strings.
$F$ is not an ordinary vertex operator; for example, it does not appear to be well defined for generic $Q$,
and the size of its condensate diverges for $Q$'s that amount to integer $k$'s.
We show that these problems are resolved by thinking about $F$ as the fusion of $W^+$ and $W^-$; schematically,
\eqn\master{F\sim W^+ * W^-.}
We are thus led to suggest that, while in the cigar geometry the non-perturbative corrections are described in terms of $W^{\pm}$,
when we analytically continue to the BH geometry, $W^+$ is combined with a $W^-$ to form an $F$ (see figure 1),
which describes the non-perturbative $\alpha'$ corrections in the BH geometry.
This is consistent with the fact that $F$ is mutually local with standard vertex operators in the BH geometry.

\ifig\loc{The Hartle-Hawking wave function and the non-perturbative $\alpha'$ corrections. In the Euclidean section there are strings with winding number $\pm 1$, denoted in the text by $W^{\pm}$. They  are combined in the Lorentzian section to form a folded string, $F$, that can be viewed as a fusion of $W^+$ and $W^-$.}
{\epsfxsize4.0in\epsfbox{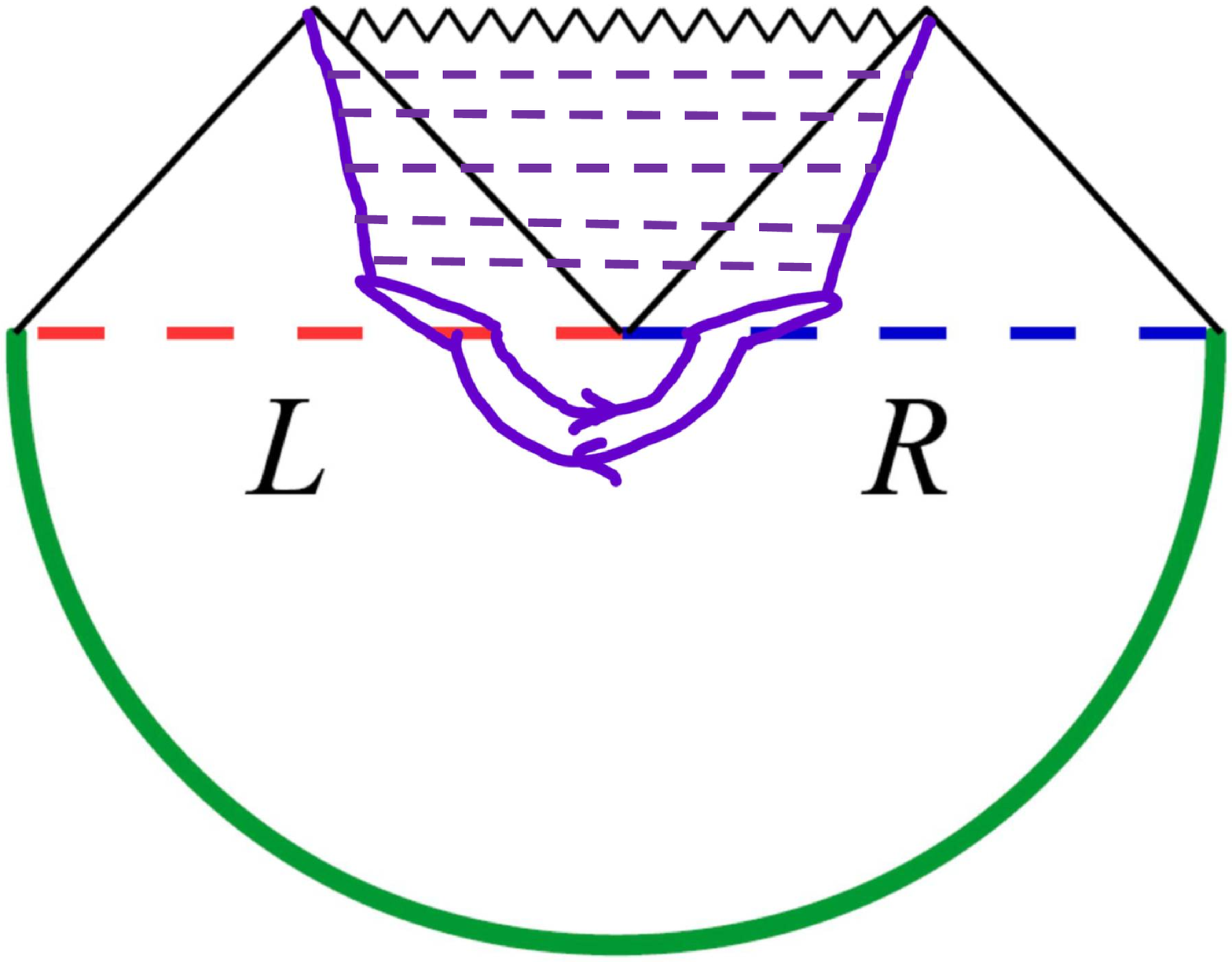}}

The rest of this note is organized as follows.
In section 2, we discuss $W^{\pm}$ and $F$ in $AdS_3$,
and in section 3, we focus on the $SL(2,\IR)/U(1)$ cigar.
In particular, we show that $F$ is the fusion of $W^+$ and $W^-$,
and sharpen the relation between them both from the CFT and space-time points of view.
Moreover, we propose that while $W^{+}$ is the wave function of the string condensate, and $W^{-}$ its conjugate,
their fusion $F$ can be viewed as the probability density.
Section 4 is devoted to the $SL(2,\IR)/U(1)$ BH.
There, we emphasize the challenge in describing the non-perturbative $\alpha'$ effects using $W^{\pm}$,
and argue that $F$ should be identified with an interior filling folded string.
In section 5, we estimate the folded string radiation,
and find it to be thermal with Hawking's temperature.

\newsec{ $AdS_3$ and (1.3)}

The operators $W^\pm$ in the $SL(2,\IR)/U(1)$ coset CFT, \aaaa,\bbbb,
are reduced from their underlying $AdS_3$ parents, which we denote by $W^\pm$ as well,
in a standard way; below, we briefly review (for details see e.g. \GiveonDXE)
some properties required to show the precise meaning of their fusion to $F$, \master.
Consider the sigma-model on $AdS_3$, in Poincar\'e coordinates,
\eqn\act{
S=\frac{k}{8\pi} \int d^2 z(\partial_{+} \phi \partial_{-} \phi - e^{2\phi} \partial_{+} \gamma_{-} \partial_{-} \gamma_{+} )~,
}
where $\gamma_{\pm}=t\pm x$ and $z_\pm=\tau\pm\sigma$ are light-cone coordinates on the boundary of $AdS_3$ and on the worldsheet, respectively, $\phi$ is the radial coordinate on $AdS_3$, and $k$ is the level of the left and right-moving $SL(2,\IR)$ current algebras.~\foot{In the superstring, the total level of $SL(2,\IR)$, $k$, receives a contribution of $k_b=k+2$ from a bosonic sigma model of the form \act, and an additional contribution of $-2$ from three free fermions in the adjoint representation of $SL(2,\IR)$; in this paper, we work in the bosonic string,
but parameterize the level by $k=k_b-2$.}
To describe $W^{\pm}$ and $F$, it is convenient to use the Wakimoto representation \WakimotoGF, which amounts to adding auxiliary fields, $\beta_{\pm}$, and replacing \act\ by
\eqn\action{
S= \frac{1}{4\pi}\int d^2 z\left( \partial_{+} \phi \partial_{-} \phi - Q{\hat R} \phi + \beta_{-} \partial_{+} \gamma_{-}
+ \beta_{+} \partial_{-} \gamma_{+} +\lambda_I I \right)~,
}
with
\eqn\idef{I=\beta_{-}\beta_{+}e^{-Q\phi}~,\qquad Q=\sqrt{2\over k}~;}
integrating out $\beta_\pm$ in \action\ (and rescaling the remaining fields) one gets \act.
An advantage of working with \action\ is e.g. that near the boundary, $\phi\to\infty$,
the theory becomes a free CFT that describes a linear dilaton and a $\beta-\gamma$ system.
The last term in \action\ can be viewed as a marginal perturbation,
which is a singlet of the $SL(2,\IR)_L\times SL(2,\IR)_R$ current algebra;
a.k.a. the operator $I$ is a screening operator. There are three other screening operators,
$W^{\pm}$ and $F$.
Next, we first describe $F$.

In terms of the Wakimoto variables, $F$ is
\eqn\deffff{F=\left( \beta_{+} \beta_{-}\right)^{k}e^{-{2\over Q}\phi}~.}
One can verify \GiribetFY\ that this $(1,1)$ operator is a singlet of the $SL(2,\IR)_L\times SL(2,\IR)_R$ current algebra,
and thus it should be added to the interaction Lagrangian \action,
\eqn\intform{\CL_{\rm int}=\lambda_I I+\lambda_F F~.}
The relation between $\lambda_F$ and $\lambda$ was fixed in \GiribetFT,
by calculating residues of poles that (for generic $k$) cannot be obtained using $I$,
\eqn\gn{\pi\lambda_F\Delta(k)=\left(\pi\lambda_I\Delta(1/k)\right)^{k},~~~~~~\Delta(x)\equiv{\Gamma(x)\over\Gamma(1-x)}~.}
Note that, at least for non-integer $k$, $F$ is not well defined, since it contains non-integer powers of the fields.
The calculations in \GiribetFT\ were done via a continuation from integer $k$'s;
we shall get back to this issue momentarily.

Now, we turn to the $W^{\pm}$ parents;
in terms of their $SL(2,\IR)$ quantum numbers, $j;m,\bar m$,
they take the form (see e.g. \GiveonDXE\ for details)
\eqn\phiphi{
W^{\pm}\equiv \Phi^{\omega=\pm 1}_{j={k\over 2}-1;m=\bar m=\pm{k+2\over 2}}~,}
where $\omega$ is the winding around the boundary circle.
Since they, as well as $I$ and $F$, commute with the $SL(2,\IR)_L\times SL(2,\IR)_R$ currents, they survive in $SL(2,\IR)/U(1)$ as is.
Hence, these $W^\pm$ are nothing but those denoted by $W^\pm$ in \aaaa, with $\beta=2\pi\sqrt{(k+2)\alpha'}$,~\foot{In
the superstring, $\beta=2\pi\sqrt{k\alpha'}$.}
except there we viewed them as operators in the coset CFT,
and here we pointed out that they exist already in the underlying CFT on $AdS_3$.
In particular, we can thus read off their behavior at large $\phi$ from \aaaa;
it is $W^\pm\sim\exp(-\phi/Q)$, in harmony with \master\ and \deffff.

The above discussion implies \GiveonDXE\ also that $W^\pm$ in \phiphi\ are the generalized FZZ duals of $I$.
Hence, since $I$ condenses, they must condense too, a.k.a.
\eqn\newintform{\CL_{\rm int}=\lambda_I I+\lambda_W\left(W^++W^-\right)~.}
This may appear to be in tension with \intform, but since, as we show in the next section,
$F$ is an integral over the OPE of $W^+$ and $W^-$,
there is no need to add to \newintform\ the $\lambda_F F$ condensate of \intform.
This leads us to conjecture a duality, that at first seems strange, between \newintform\ and \intform.
The operational meaning of the duality is that the non-perturbative $\alpha'$ corrections can be described
either in terms of \newintform\ or in terms of  \intform.

For this duality to hold, it must be that, inside correlation functions in $AdS_3$,
\eqn\replacef{\lambda^2_W \int d^2z W^+(z)W^-(w)=  C \lambda_F F(w)~,}
with a $C$ that is compatible with \gn\ and other known results \GiveonUP.
Since $W^{\pm}$ and $F$ (as well as $I$) are singlets of the $SL(2,\IR)_L\times SL(2,\IR)_R$ current algebra,
\replacef\ (as well as \intform\ and \newintform) should hold for any coset or orbifold of $AdS_3$.
In the next section, evidence for the duality is provided
in the $SL(2,\IR)/U(1)$ coset CFT describing the Euclidean cigar,
where some aspects of the calculations are easier.
Here, we comment that, if correct, \replacef\ clarifies the puzzle concerning $F$: the operator
$F$ is not well defined for generic $k$, since it is not a fundamental field,
but rather a composite operator, that should be viewed as a bound state (in the worldsheet)
of $W^+$ and $W^-$, which are well defined operators for generic $k$.

\newsec{The cigar and (1.3)}

In this section, we inspect the cigar $SL(2,\IR)/U(1)$ coset CFT, and  discuss \master\ in more details.
In the first subsection, we focus on the CFT point of view, in particular, on \replacef.
In the second subsection, we consider some target-space aspects of \master.

\subsec{CFT point of view}

In the previous section, we argued that \master\ has a precise meaning:
$F$ should be viewed as a bound state of $W^+$ and $W^-$ in the sense of \replacef.
This is a strong claim: it implies that in correlation functions
we can replace integrals over pairs of $W^+$ and $W^-$ with $F$'s, e.g. that
\eqn\bulkcorr{\lambda_W^{2n}\left\langle \prod_{j=1}^n\int d^2 z_jW^+(z_j)\prod_{l=1}^n \int d^2w_lW^-(w_l)\cdots\right\rangle
=(C\lambda_F)^n\left\langle \prod_{l=1}^n \int d^2w_lF(w_l)\cdots\right\rangle}
Taking into account the combinatoric factor, which on the l.h.s. of \bulkcorr, obtained by expanding
$\langle e^{\int d^2z{\cal L}_{\rm int}}\cdots\rangle$ with ${\cal L}_{\rm int}$ in \newintform, is
${1\over (2n)!}{2n\choose n}$=${1\over(n!)^2}$,
whereas in the calculation based on \intform\ on the r.h.s. is just ${1\over n!}$,
\bulkcorr\ implies that the integrals over $z_j$ on its l.h.s. must only receive contributions from regions
where each of the $W^+$'s approaches one of the $W^-$'s (\ie\ one of the $z_j$ is close to one of the $w_l$),
and the contribution from one such region is given by the r.h.s. of \bulkcorr.
This is the essence of the conjectured duality between \newintform\ and \intform.
In  fact, it was already shown in \GiribetFT, in some non-trivial cases,
that if one replaces the Lagrangian \newintform\ by \intform,
and focuses on the coefficient of $\lambda_F^n$, instead of $\lambda_W^{2n}$ in \bulkcorr,
one gets correct results (if \gn\ is satisfied).

Next we calculate $C$, and verify that its value is self-consistent with the above duality.
For this purpose, it is sufficient to look at
\eqn\expcor{(\lambda_W)^2\int d^2z\int d^2w\langle W^+(z) W^-(w)\cdots\rangle~.
}
To find the contribution to the integral over $z$ from the region where $z$ is close to $w$, we work with a Euclidean worldsheet,
and use the free field OPE of the operators \aaaa,
\eqn\opeww{W^+(z)W^-(w)={1\over |z-w|^{2(k+1)}}e^{i\sqrt{k+2\over2}\left[(x_L-x_R)(z)-(x_L-x_R)(w)\right]-{1\over Q}(\phi(z)+\phi(w))}~,
}
where we set $\alpha'=2$.
To focus on the contribution of the region $z\to w$, we expand the exponential in $z-w$. Keeping only the first non-trivial term, we have
\eqn\openwwapprox{W^+(z)W^-(w)\simeq{1\over |z-w|^{2(k+1)}} \left|
e^{-(z-w)\left({1\over Q}\partial\phi-i\sqrt{k+2\over2}\partial x\right) }\right|^2
e^{-{2\over Q}\phi(w)}~,
}
where $\simeq$ in \openwwapprox\ stands for the fact that we neglected higher powers of $z-w$ and it's c.c. in the exponential.
Plugging \openwwapprox\ into \expcor, and using the identity
\eqn\identity{\int d^2z |z|^{2\alpha} e^{ip\cdot z}=\pi\Delta(\alpha+1)\left|{p\over2}\right|^{-2(\alpha+1)}~,
}
with $\alpha=-k-1$, the integral over $z$ in \expcor\ gives~\foot{It is curious
that a similar consideration, referred to as a ``curious result,"
appears in \PolchinskiTA, while considering other aspects of BH physics.}
\eqn\expcornew{\pi\Delta(-k)(\lambda_W)^2\int d^2w\left\langle \left|{1\over Q}\partial\phi-i\sqrt{k+2\over2}\partial x\right|^{2k}e^{-{2\over Q}\phi(w)}
\cdots\right\rangle~.}
The operator inside the correlation function \expcornew\ can be recognized using the results of \BershadskyIN\ to be the leading contribution to $F$ \deffff.  Hence,~\foot{The fact that in \expcornew\ we only get the leading term in the expansion of $F$ is due to the fact that we truncated the expansion of \opeww\ at first order. The higher order terms in $F$ come from higher order terms in that expansion.}
\eqn\lllll{\lambda_F=-\pi\Delta(-k)(\lambda_W)^2~.}
One can combine \lllll\ with \gn\ to find a relation between $\lambda_W$ and $\lambda_I$,
\eqn\dldl{{\pi\lambda_W\over k}=\left(\pi\lambda_I\Delta(1/k)\right)^{k\over2}~,}
which agrees~\foot{Up to a sign, which is due to a different convention for the couplings.} with  \GiveonUP.

To recapitulate, note that the authors of \GiribetFT\ actually showed that the calculation based on the r.h.s. of \bulkcorr\ gives the correct result for the special cases of the two and three point functions. Our proposed duality implies that their results should generalize to arbitrary  correlation functions.
Moreover, the discussion  above does {\it not} explain the duality, that is how come the l.h.s. of \bulkcorr\ and its r.h.s. agree. The phenomenon here is reminiscent of confinement, with the winding operators $W^\pm$ playing the roles of quarks and anti-quarks, and $F$ playing the role of a meson. Needless to say that it would be very helpful to understand the mechanism behind this ``confinement."
A partial explanation  is provided by a Gross-Mende like reasoning \refs{\GrossKZA,\GrossAR}.
Consider, for example, the l.h.s. of equation \replacef.
Equation \aaaa\  implies  that there is an  attractive potential on the worldsheet, $V\sim k \log|z-w|^{2}$, between $W^{+}(z)$ and $W^{-}(w)$;
it is likely that this is part of the explanation for the formation of $F$.

\subsec{Space-time point of view}

In this subsection, we discuss the space-time interpretation of $W^{\pm}$, $F$ and the relation between them.
Here, we consider the cigar geometry, but this will be proved useful also in the BH case, discussed in the next sections.

Let us first review \KutasovRR\ the effective space-time action that determines the $\phi$ dependence of $W^{\pm}$.
Consider a string that winds the Euclidean circle in the cigar geometry,
\eqn\metaa{ds^2=2k \tanh^2(\phi/\sqrt{2k})d\theta^2+d\phi^2,\qquad \theta\sim\theta+2\pi,\qquad
e^{2\Phi}={g_s^2 \over \cosh^2(\phi/\sqrt{2k})}~.}
We present the fermionic case, and set $\alpha'=2$.
Its effective space-time action reads \refs{\KutasovRR,\GiveonICA}
\eqn\actionw{S(\omega=\pm 1)={1 \over 2}\int_0^{\infty} d\phi\sqrt{k/2}\sinh(\sqrt{2/k}\phi)\left( (\partial_{\phi}\Psi_{\pm})^2+M^2(\phi)\Psi_{\pm}^2\right)~,}
where
\eqn\massw{M^2(\phi)=-1 +{k\over 2}\tanh^2(\phi/\sqrt{2k})~.}
The $-1$ comes
from the fact that a string that winds the Euclidean time direction is a tachyonic mode (see e.g. \AtickSI) and the second term is the energy associated with the winding of the string.
This action admits a  zero energy solution  \GiveonICA,
\eqn\solw{\Psi^{\pm}(\phi)=\cosh^{-k}(\phi/\sqrt{2k})~.}
At large $\phi$, it behaves like $\exp(-\phi/Q)$, as it should, \aaaa.
This approach, however, reveals the profile of $W^{\pm}$ at smaller $\phi$.
In particular, we get
$\Psi^{\pm}\sim \exp(-\phi^2 /4)$
for  $\phi\ll \sqrt{k}$. It is natural to interpret $\Psi^{\pm}(\phi)$ as the wave function of the string  condensate in the cigar.

Support for this interpretation comes from the following consideration, that goes back to \WittenZW. There, a temporal Wilson loop \refs{\PolyakovVU,\SusskindUP} at the boundary was identified in the bulk with a string that wraps the cigar. In a closely related setup,
this approach was used to determine the analog of $\lambda_W$ \AharonyAN. Here, we point out that the same approach can be used to determine,
in the spirit of \HartleAI, $\Psi^{\pm}(\phi)$.
The saddle point approximation of $\Psi^{\pm}(\phi)$ is
\eqn\pathin{\Psi^{\pm}(\phi)\sim \exp(-S_W^{\pm}(\phi))~,}
where
$S_W^{\pm}(\phi)$ is the Nambu-Goto action of a string that wraps the cigar from the tip to $\phi$,
\eqn\acts{S^{\pm}_W(\phi)=\sqrt{k/2} \int_0^{\phi} d\rho \tanh(\rho/\sqrt{2k})~.}
Indeed,
\eqn\popo{\exp(-S^{\pm}_W(\phi))=\cosh^{-k}(\phi/\sqrt{2 k})~,}
in agreement with \solw.
In the above, $\Psi^+=\Psi^-$ and thus the wave function is real. This, however, is not quite the case.
A $B$-field will contribute a phase $\exp\left( \pm i\int_0^{\phi} B\right)$
to $\Psi^{\pm}$, as expected from a wave function of a string that winds around a circle.
The sign dependence comes from the orientation of the string that wraps the cigar.
Since the cigar is two dimensional, it contains no $H$ field and so we can set $B=0$.
However, one may also consider a gauge transformation, $B=dA$,
where $A$ is a one-form gauge parameter, in which case it is the phase of $\Psi^\pm(\phi)$,~\foot{This
gauge redundancy is related to the freedom to add an arbitrary
constant to the argument of the cosine in the Sine-Liouville potential $\lambda_W(W^++W^-)$, \aaaa.}
a.k.a.~\foot{In \WittenZW, the phase of the wave function of the string
played an important role in determining when the temporal Wilson line condenses (see also \AharonyQU).}
\eqn\pass{\Psi^\pm(\phi)=\cosh^{-k}(\phi/\sqrt{2 k})\,\exp\left(\pm i \oint A (\phi)\right).}

There seems to be a natural interpretation to $F$ as well; its $\phi$ profile is the square of the profile of $W$ (see e.g. \deffff).
In terms of a string configuration, this is due to the fact that $F$ is identified with a folded string that goes all the way from the tip to   $\phi$ and back, while wrapping the Euclidean time direction. This configuration, that does not solve the equations of motion at the fold, is closely related to the worldsheet instantons of \MaldacenaKM\ (that also do not solve the equations of motion), and its action is twice that of \acts. Hence, the $\phi$ profile of $F$ is
\eqn\popo{\exp(-S_F(\phi))=\cosh^{-2k}(\phi/\sqrt{2 k})~,}
which agrees with \deffff\ at large $\phi$.

The profile of $F$ does not contain a phase, since the configuration associated with it has no boundary.
This, and the discussion above, suggest that the $\phi$ profile of $F$ may be interpreted as the probability density, $\rho(\phi)$.
From this point of view, the interpretation of \master\ is simply that
\eqn\density{\rho(\phi)=|\Psi(\phi)|^2~.}
These considerations suggest that $F$ and $W^{\pm}$ describe the same object -- the string condensate. $W^{\pm}$ describe its wave function while $F$ describes its probability density. At first, this seems to be in tension with claims made in the previous section. There, we argued that $F$
can replace $W^{\pm}$, which implies that it contains the same information. Normally, due to the phase, the wave function contains more information than the probability density. Here, however, the phase contains no information, as it can be gauged away.

\newsec{Lorentzian black hole}

The discussion in the previous sections implies that in the $SL(2,\IR)/U(1)$ cigar,
$F$ describes the same non-perturbative $\alpha'$ corrections as $W^{\pm}$.
Still, since $F$ can be viewed as a ``bound state" of $W^{+}$ and $W^{-}$,
it seems to be less fundamental than $W^{\pm}$.
In this section, we argue that, in a certain sense, the opposite is happening in the $SL(2,\IR)/U(1)$ BH theory.
Working with $F$ seems to be straightforward, while working with $W^{\pm}$ is challenging.

Let us start by describing the subtleties in working with $W^{\pm}$.
For this, it is sufficient to consider the region far from the BH.
There, the background is described by a time direction $t$ times a linear-dilaton direction $\phi$.
Standard states in this background carry energy, $E$, and are represented by vertex operators, $V_E$, which behave like
\eqn\ver{V_E\sim \exp(-iE(t_L+t_R)),}
where $t_{L,R}$ are the left and right-handed modes of $t=t_L+t_R$ on the worlsheet.
The problem with $W^{\pm}$, which in the Lorentzian BH take the form \bbbb,
is that they are not mutually local with $V_E$.
In fact, they are not mutually local in an intriguing way; when $W^{\pm}$ goes around $V_E$ it picks up a Boltzmann factor $\exp(\pm \beta E)$.
This reflects the thermal nature of the BH, that is manifest by the fact that $t$ is periodic in the imaginary direction,
$t\sim t+i \beta.$
Namely, when $W^{\pm}$ goes around $V_E$ on the worldsheet, it shifts $t$ in the imaginary direction, $t\to t\pm i\beta$, in space-time.
As far as we can see, at the moment,
this observation does not help in utilizing a direct prescription for well defined string theory calculations.
The problem remains: typical correlators take the form
\eqn\amp{\left\langle \prod_{j=1}^n\int d^2 z_jW^+(z_j)\prod_{l=1}^n \int d^2w_lW^-(w_l) \int d^2v V_E (v) \cdots\right\rangle,}
and the fact that $W^{\pm}$ \bbbb\ are not mutually local with $V_E$ \ver\ implies that the integrals over the $z's$ and $w's$
are not well defined.~\foot{It is reasonable that this  issue, which appears already in $AdS_3$ when working in the hyperbolic $J^2$ basis,
is related to the fact that we are considering a non-trivial Lorentzian space-time
-- the quotient of $SL(2,\IR)$ in the $J^2$ direction, while embedding in it a Euclidean worldsheet.}

Working with $F$, on the other hand, we do not encounter this subtlety.
Since $F$  carries no winding, it is mutually local with $V_E$.
Therefore, correlators of the form
\eqn\bff{\left\langle \prod_{l=1}^n \int d^2w_lF(w_l) \int dv^2 V_E(v) \cdots\right\rangle}
do not involve the issue discussed above.
In other words, via the analytic continuation of \bulkcorr\ in space-time,
its l.h.s. includes correlators of the type \amp, that we cannot calculate directly,
while the r.h.s. includes correlatos of the type \bff, that we can calculate.
We are thus led to define the l.h.s. of the analog of \bulkcorr\ in the BH
by the continuation of its r.h.s. to Lorentzian space-time.
This suggests that $F$ should have a clear and simple meaning in the BH case.
Indeed, the discussion at the end of the previous section implies that $F$
should be identified with the folded string that fills the entire BH, discussed in \ItzhakiGLF.

To make this precise, it is helpful to use the Hartle-Hawking (HH) wave function \HartleAI\ approach,
in which the initial condition for the  state in the Lorentzian section is determined by the Euclidean section.
As described in the previous section, in the cigar, the space-time interpretation of $F$ is of a folded string instanton,
that goes all the way from the tip to a certain $\phi$ and back, while wrapping the Euclidean time direction.
The HH procedure instructs us to cut the cigar into two halves and analytically continue the upper half into the Lorentzian section.
The analytic continuation does not involve $\phi$, so it does not affect \popo.
We thus interpret  \popo\ as  the probability for the folded string to escape a distance $\phi$ outside the BH.

This goes well with \ItzhakiGLF.
Classically, a string that is outside the $SL(2,\IR)/U(1)$ BH cannot fold towards the BH \refs{\MaldacenaHI,\ItzhakiGLF}.
This is the reason why the classical folded string lives entirely inside the BH.
Perturbative corrections in $\alpha'$ cannot change this.
Equation \popo\ implies that non-perturbative effects in $\alpha'$ can.
According to  \popo, as we increase $\phi$ we decrease, exponentially fast, the probability to find a folded string.

A couple of comments are in order:
\item{(1)} In \ItzhakiGLF, it was argued that
           the folded string that condenses does not break the symmetries of the BH geometry.
           In particular, its energy vanishes.
           This is in agreement with the fact that the operator $F$ is a singlet of $SL(2,\IR)_R\times SL(2,\IR)_L$.
\item{(2)} As was argued above, the condensation of $F$ appears already on the sphere. Thus, we expect the number of folded strings to scale like $1/g_s^2$. This is a necessary (but not a sufficient) condition  for the  conjecture made in \AttaliGOQ, that the back reaction of the folded string cloaks  the BH.

\noindent

To recapitulate, the picture that emerges from this, and the previous section, is that
in the Euclidean section, the non-perturbative $\alpha'$ corrections are described more naturally in terms of $W^{\pm}$,
while in the Lorentzian section, in terms of $F$. This suggests that a stringy way to present the HH wave function,
taking into account the non-perturbative corrections in $\alpha'$, is figure 1.

\newsec{Folded string radiation}

In this section, we take advantage of the fact that $F$ describes a folded string,
to estimate the folded string radiation. We show that it is thermal, with the same temperature as the Hawking temperature.

As discussed above and in \ItzhakiGLF, classically, the folded string is leaving inside the BH.
Outside the BH, the folded string has a profile  \popo, that describes how it tunnels into the classically forbidden region outside the BH. The fact that the folded string has a tail outside the BH implies that a piece of it can break away and radiate to infinity.
For this to happen, the folded string has to break at some point, $\phi_B$, outside the BH. The probability for a folded string that is stretched to $\phi=\phi_0$ to break at $\phi_B$  goes like
\eqn\breakk{P(\phi_B, \phi_0)\sim g_s^2 \cosh^{-2k}(\phi_0/\sqrt{2 k})\cosh^{-2k}(\phi_B/\sqrt{2 k})~.}
The factor of $g_s^2$ is standard in string theory. The factor of
$\cosh^{-2k}(\phi_0/\sqrt{2 k}) $ is the probability  to have an initial state of a folded string that is stretched all the way to $\phi_0$. When the folded string breaks, one side of it can fold classically towards infinity \refs{\MaldacenaHI,\ItzhakiGLF}.
Hence, this folding does not lead to an extra suppression.
However, the other side needs to fold towards the BH. This is forbidden classically, but, as discussed above,
can happen non-perturbatively in $\alpha'$. This leads to the factor of $\cosh^{-2k}(\phi_B/\sqrt{2 k})$ in \breakk.
Given an initial state of a folded string, that is stretched all the way to $\phi_0$,
it is clear that in order to maximize \breakk, we should take small $\phi_B$;
that is the string breaks just outside the horizon. In that case, we have
\eqn\breakkk{P(\phi_0)\sim g_s^2 \cosh^{-2k}(\phi_0/\sqrt{2 k})~.}
This quantity behaves like $\exp(-\beta E)$, where $\beta$ is the inverse Hawking temperature, and $E$ is the energy of the
piece of the broken folded string that can now escape to infinity,
\eqn\energyfol{E={2 \over 4\pi}\int_0^{\phi_0}d\phi \sqrt{g_{00}}={1\over 2\pi} \int_0^{\phi_0} d\phi \tanh(\phi/\sqrt{2k})~,}
where again we are working with $\alpha'=2$.

Some comments:
\item{(1)} This calculation is the Lorentzian analog of the Euclidean calculation presented in subsection 3.2.
\item{(2)} The discussion here is schematic. For example, it is done in $2d$, where there are no propagating degrees of freedom;
a more precise analysis must involve polarization in the other directions.
\item{(3)} Note that \breakkk\ scales like $g_s^2$. Since, as argued before, the number of folded strings scales like $1/g_s^2$, the total energy flux does not depend on $g_s$ (as it should).

\noindent

The main point here is that, unlike in Hawking's calculation \HawkingSW, where nothing (or more precisely the BH vacuum) is burning,
here something -- the folded string -- is burning at the same temperature. We have seen this before many times in string theory
(for reviews, see e.g. \refs{\MaldacenaKY,\PeetHN}), where by going to weak coupling,
the BH radiation is replaced by a system of burning D-branes.
The novelty here is that we see the burning objects on the gravity side.

\bigskip
\noindent{\bf Acknowledgements:}
We thank D. Kutasov  for collaboration on many parts of this paper. This work is supported in part
by a center of excellence supported by the Israel Science Foundation (grant number 2289/18).

\listrefs

\end

**** Entropy ****

\noindent
\newsec{List of things we wish to add}

1. Polynomials.

2. Bound state in GR vs. GFZZ $(W_l)$.

3.  $F_l$ (file k-l).

4. Class solution: W's and F.


\lref\SugawaraHMA{
  Y.~Sugawara,
  ``“Analytic continuation” of $N=2$ minimal model,''
PTEP {\bf 2014}, no. 4, 043B02 (2014).
[arXiv:1311.4708 [hep-th]].
}

\lref\GIKthree{A.~Giveon, N.~Itzhaki and D.~Kutasov, to appear.}

\lref\BekensteinUR{
  J.~D.~Bekenstein,
  ``Black holes and entropy,''
Phys.\ Rev.\ D {\bf 7}, 2333 (1973).
}

\lref\HawkingSW{
  S.~W.~Hawking,
  ``Particle Creation by Black Holes,''
Commun.\ Math.\ Phys.\  {\bf 43}, 199 (1975), [Erratum-ibid.\  {\bf 46}, 206 (1976)].
}

\lref\MaldacenaHW{
  J.~M.~Maldacena and H.~Ooguri,
  ``Strings in AdS(3) and SL(2,R) WZW model 1.: The Spectrum,''
J.\ Math.\ Phys.\  {\bf 42}, 2929 (2001).
[hep-th/0001053].
}

\lref\PolchinskiRR{
  J.~Polchinski,
  ``String theory. Vol. 2: Superstring theory and beyond,'' section 10.4.
}

\lref\BarbonNW{
  J.~L.~F.~Barbon and E.~Rabinovici,
  ``Remarks on black hole instabilities and closed string tachyons,''
Found.\ Phys.\  {\bf 33}, 145 (2003).
[hep-th/0211212].
}

\lref\ArgurioTB{
  R.~Argurio, A.~Giveon and A.~Shomer,
  ``Superstrings on AdS(3) and symmetric products,''
JHEP {\bf 0012}, 003 (2000).
[hep-th/0009242].
}

\lref\GiveonWN{
  A.~Giveon, A.~Konechny, A.~Pakman and A.~Sever,
  ``Type 0 strings in a 2-d black hole,''
JHEP {\bf 0310}, 025 (2003).
[hep-th/0309056].
}

\lref\KutasovRR{
  D.~Kutasov,
  ``Accelerating branes and the string/black hole transition,''
[hep-th/0509170].
}

\lref\fzz{
V.A. Fateev, A.B.
Zamolodchikov and Al.B. Zamolodchikov, unpublished.
}

\lref\KazakovPM{
  V.~Kazakov, I.~K.~Kostov and D.~Kutasov,
  ``A Matrix model for the two-dimensional black hole,''
Nucl.\ Phys.\ B {\bf 622}, 141 (2002).
[hep-th/0101011].
}

\lref\GiveonPX{
  A.~Giveon and D.~Kutasov,
  ``Little string theory in a double scaling limit,''
JHEP {\bf 9910}, 034 (1999).
[hep-th/9909110].
}


\lref\GiveonKP{
  A.~Giveon and N.~Itzhaki,
  ``String Theory Versus Black Hole Complementarity,''
JHEP {\bf 1212}, 094 (2012).
[arXiv:1208.3930 [hep-th]].
}

\lref\GiveonICA{
  A.~Giveon and N.~Itzhaki,
  ``String theory at the tip of the cigar,''
JHEP {\bf 1309}, 079 (2013).
[arXiv:1305.4799 [hep-th]].
}

\lref\MertensPZA{
  T.~G.~Mertens, H.~Verschelde and V.~I.~Zakharov,
  ``Near-Hagedorn Thermodynamics and Random Walks: a General Formalism in Curved Backgrounds,''
JHEP {\bf 1402}, 127 (2014).
[arXiv:1305.7443 [hep-th]].
}

\lref\MertensZYA{
  T.~G.~Mertens, H.~Verschelde and V.~I.~Zakharov,
  ``Random Walks in Rindler Spacetime and String Theory at the Tip of the Cigar,''
JHEP {\bf 1403}, 086 (2014).
[arXiv:1307.3491 [hep-th]].
}

\lref\GiveonHSA{
  A.~Giveon, N.~Itzhaki and J.~Troost,
  ``The Black Hole Interior and a Curious Sum Rule,''
JHEP {\bf 1403}, 063 (2014).
[arXiv:1311.5189 [hep-th]].
}

\lref\GiveonHFA{
  A.~Giveon, N.~Itzhaki and J.~Troost,
  ``Lessons on Black Holes from the Elliptic Genus,''
JHEP {\bf 1404}, 160 (2014).
[arXiv:1401.3104 [hep-th]].
}

\lref\MertensNCA{
  T.~G.~Mertens, H.~Verschelde and V.~I.~Zakharov,
  ``The thermal scalar and random walks in $AdS_3$ and $BTZ$,''
JHEP {\bf 1406}, 156 (2014).
[arXiv:1402.2808 [hep-th]].
}

\lref\MertensCIA{
  T.~G.~Mertens, H.~Verschelde and V.~I.~Zakharov,
  ``Near-Hagedorn Thermodynamics and Random Walks - Extensions and Examples,''
JHEP {\bf 1411}, 107 (2014).
[arXiv:1408.6999 [hep-th]].
}

\lref\WakimotoGF{
  M.~Wakimoto,
  ``Fock representations of the affine lie algebra A1(1),''
Commun.\ Math.\ Phys.\  {\bf 104}, 605 (1986).
}

\lref\BernardIY{
  D.~Bernard and G.~Felder,
  ``Fock Representations and BRST Cohomology in SL(2) Current Algebra,''
Commun.\ Math.\ Phys.\  {\bf 127}, 145 (1990).
}

\lref\GiveonNS{
  A.~Giveon, D.~Kutasov and N.~Seiberg,
  ``Comments on string theory on AdS(3),''
Adv.\ Theor.\ Math.\ Phys.\  {\bf 2}, 733 (1998).
[hep-th/9806194].
}

\lref\MertensDIA{
  T.~G.~Mertens, H.~Verschelde and V.~I.~Zakharov,
  ``On the Relevance of the Thermal Scalar,''
JHEP {\bf 1411}, 157 (2014).
[arXiv:1408.7012 [hep-th]].
}

\lref\MertensSAA{
  T.~G.~Mertens, H.~Verschelde and V.~I.~Zakharov,
JHEP {\bf 1506}, 167 (2015).
[arXiv:1410.8009 [hep-th]].
}

\lref\GiveonCMA{
  A.~Giveon, N.~Itzhaki and D.~Kutasov,
  ``Stringy Horizons,''
JHEP {\bf 1506}, 064 (2015).
[arXiv:1502.03633 [hep-th]].
}

\lref\GiribetKCA{
  G.~Giribet and A.~Ranjbar,
  ``Screening Stringy Horizons,''
Eur.\ Phys.\ J.\ C {\bf 75}, no. 10, 490 (2015).
[arXiv:1504.05044 [hep-th]].
}

\lref\MertensHIA{
  T.~G.~Mertens, H.~Verschelde and V.~I.~Zakharov,
  ``The long string at the stretched horizon and the entropy of large non-extremal black holes,''
JHEP {\bf 1602}, 041 (2016).
[arXiv:1505.04025 [hep-th]].
}

\lref\BenIsraelMDA{
  R.~Ben-Israel, A.~Giveon, N.~Itzhaki and L.~Liram,
  ``Stringy Horizons and UV/IR Mixing,''
JHEP {\bf 1511}, 164 (2015).
[arXiv:1506.07323 [hep-th]].
}

\lref\MertensOLA{
  T.~G.~Mertens,
  ``Hagedorn String Thermodynamics in Curved Spacetimes and near Black Hole Horizons,''
[arXiv:1506.07798 [hep-th]].
}

\lref\LinLFA{
  J.~Lin,
  ``Bulk Locality from Entanglement in Gauge/Gravity Duality,''
[arXiv:1510.02367 [hep-th]].
}

\lref\BenIsraelETG{
  R.~Ben-Israel, A.~Giveon, N.~Itzhaki and L.~Liram,
  ``On the Stringy Hartle-Hawking State,''
JHEP {\bf 1603}, 019 (2016).
[arXiv:1512.01554 [hep-th]].
}

\lref\CottrellNSU{
  W.~Cottrell and A.~Hashimoto,
  ``Resolved gravity duals of ${\cal N}=4$ quiver field theories in 2+1 dimensions,''
[arXiv:1602.04765 [hep-th]].
}



\lref\ElitzurCB{
  S.~Elitzur, A.~Forge and E.~Rabinovici,
  ``Some global aspects of string compactifications,''
Nucl.\ Phys.\ B {\bf 359}, 581 (1991).
}

\lref\MandalTZ{
  G.~Mandal, A.~M.~Sengupta and S.~R.~Wadia,
  ``Classical solutions of two-dimensional string theory,''
Mod.\ Phys.\ Lett.\ A {\bf 6}, 1685 (1991).
}

\lref\WittenYR{
  E.~Witten,
  ``On string theory and black holes,''
Phys.\ Rev.\ D {\bf 44}, 314 (1991).
}

\lref\DijkgraafBA{
  R.~Dijkgraaf, H.~L.~Verlinde and E.~P.~Verlinde,
  ``String propagation in a black hole geometry,''
Nucl.\ Phys.\ B {\bf 371}, 269 (1992).
}


\lref\KazamaQP{
  Y.~Kazama and H.~Suzuki,
 ``New N=2 Superconformal Field Theories and Superstring Compactification,''
Nucl.\ Phys.\ B {\bf 321}, 232 (1989).
}

\lref\BershadskyIN{
  M.~Bershadsky and D.~Kutasov,
  ``Comment on gauged WZW theory,''
Phys.\ Lett.\ B {\bf 266}, 345 (1991).
}

\lref\HorowitzNW{
  G.~T.~Horowitz and J.~Polchinski,
  ``A Correspondence principle for black holes and strings,''
Phys.\ Rev.\ D {\bf 55}, 6189 (1997).
[hep-th/9612146].
}

\lref\HorowitzJC{
  G.~T.~Horowitz and J.~Polchinski,
  ``Selfgravitating fundamental strings,''
Phys.\ Rev.\ D {\bf 57}, 2557 (1998).
[hep-th/9707170].
}

\lref\GiveonPR{
  A.~Giveon and D.~Kutasov,
  ``Fundamental strings and black holes,''
JHEP {\bf 0701}, 071 (2007).
[hep-th/0611062].
}